# Enhancement of Two photon absorption with Ni doping in the dilute magnetic Semiconductor ZnO Crystalline Nanorods

Amit Kumar Rana[1, 2, a)], Aneesh. J[3, a)], Yogendra Kumar[1, 2], Arjunan. M. S[1, 2], K. V. Adarsh[3], Somaditya Sen[1, 2], and Parasharam M. Shirage[1, 2,*]

[1]Department of Physics and [2] Centre of Materials Science and Engineering, Indian Institute of Technology Indore. Simrol Campus, Khandwa Road, Indore-452020, India

3 Department of Physics, Indian Institute of Science Education and Research, Bhopal-462023, India.

In this Letter, we have investigated the third-order optical nonlinearities of high-quality Ni doped ZnO nanorods crystallized in wurtzite lattice, prepared by the wet chemical method. In our experiments, we found that the two photon absorption coefficient ($\beta$) increases by as much as 14 times i.e. 7.6 ±0.4 to 112±6 cm/GW, when the Ni doping is increased from 0 to 10 %. The substantial enhancement in $\beta$ is discussed in terms of the bandgap scaling and Ni doping. Furthermore, we also show that the optical bandgap measured by UV-Vis and photoluminescence spectroscopies, continuously redshift with increasing Ni doping concentration. We envision that the strong nonlinear optical properties together with their dilute magnetic effects, they form an important class of materials for potential applications in magneto-optical and integrated optical chips.

[a)]Authors contributed equally to this work
*Author to whom correspondence should be addressed; electronic mail: pmshirage@iiti.ac.in and paras.shirage@gmail.com

Semiconductors with multifunctional properties are critical in the next-generation optoelectronic devices. Among such materials, wide direct bandgap II-VI semiconductor ZnO surge in much interest in transparent conductors [1], photo-catalyst [2], gas sensors [3], field-effect transistor [4], light emitting diodes and various optoelectronic devices [5-8]. More importantly, the optical and electrical properties of ZnO can be effectively tuned by controlling the intrinsic defects, which creates localized states in the bandgap [9, 10]. Likewise, ZnO can also be easily doped with magnetic impurities. For example, pristine ZnO doesn't show the magnetic response; however, when doped with transition metals it exhibits weak ferromagnetism [11]. Because of their potential in spintronics [12], there is a high interest in the search for magnetic semiconductors. Furthermore, antiferromagnetic, ferromagnetic and ferrimagnetic properties are also reported for ZnO by doping with Co, Fe, and Co-Fe respectively [11, 13, 14].

Recent advances in the chemical synthesis of ZnO have enabled to prepare them in complex nanocrystals with an increased degree of structural complexity and shape [15]. This opens up another degree of freedom to harness many optical phenomena that result from quantum confinement effects. Recently, two-photon absorption (TPA) induced UV emission is reported in ZnO nanowires [16] that can be used to fabricate low-cost nanolasers for photonic circuitries and sensing systems. Although, extensive studies have been carried out toward revealing the size-dependent optical and nonlinear optical properties of ZnO nanocrystals, however, there were not many studies on the variation of such effects with the doping of transition metals. This is to the fact that their linear optical properties can be tuned just by the doping. For example, various theoretical and experimental studies on the nonlinear optical absorption in semiconductors exhibit an inverse cubic relation with band gap for the two photon absorption coefficient ($\beta$). Thus, such studies are important to explore how the nonlinear optical properties of semiconductor NCs depend on the doping, but also provide insight to engineer it. In this Letter, we demonstrate that the two photon absorption coefficient (β) increases by as much as 14 times with Ni doping in high-quality ZnO nanorods crystallized in wurtzite lattice. The significant increase in β is a result of the redshift in optical bandgap with Ni doping and follows the bandgap scaling.

Ni-doped ZnO samples were synthesized by the wet chemical method. Analytical grade zinc nitrate, nickel nitrate, and ammonia were used as the raw materials for the synthesis of 100 mM solutions of pure and Ni doped ZnO nanorods. For this, first we have dissolved the stoichiometric concentrations of zinc nitrate, and nickel nitrate in 100 ml double distilled water and stirred for 30 minutes. Then under constant stirring, aqueous ammonia solution was added continuously to maintain a pH of 12.8±0.2 (measured with the help of a digital pH meter). Finally, a transparent solution is obtained in the case of pure ZnO and blue solutions for Ni doping doped ZnO nanorods. In the next step, the solutions were kept at 85 $^o$C for 2 hours and then the precipitates were separated and annealed at 200 $^o$C (for 2 hrs.). The samples with Ni doping concentration of 0, 5, and 10 % were named as ZnO0, ZnO5, and ZnO10 respectively.

The synthesized nanorods were characterized by scanning electron microscopy (SEM) images, and the crystallinity was measured by X-ray diffraction (XRD). Optical properties were measured by using UV-vis and photoluminescence spectroscopies. To estimate the third order nonlinear optical properties, we have used the conventional open aperture Z-scan [17] method, which measures the total transmittance as a function of incident laser intensity. In our Z-scan experiment, 5 ns pulses centered at 532 nm from the second harmonic of Nd-YAG laser was used to excite the sample. The repetition rate of the laser was kept at 10 Hz to avoid the heating effects. The beam was focused by a 50 cm focal length lens, and the sample was moved along the z-axis of the beam by using a computer controlled translation stage. The Rayleigh length ($z_0$) and the beam waist in our experiment were 10 mm and ~84 µm respectively. For the Z-scan measurements, all the samples were dispersed in ethanol and used a 1 mm path length cuvette.

The SEM images of the synthesized nanorods are shown in Fig. 1(a-c). The average diameter of the ZnO0, ZnO5, and ZnO10 nanorods are ~100, ~150 and ~200 nm respectively. From the SEM images, it can be seen that there is a change in the morphology of the nanorods with increased Ni concentration. Figure 1(d) shows the XRD patterns of all the samples, which indicate that these materials exhibit wurtzite lattice, and the *c*-axis is the preferential growth direction. There are no traces of impurity related peaks correspond to Ni, which shows that the doping did not change the wurtzite structure of ZnO. The high-intensity peak of (002) in the ZnO0 sample decreases with Ni doping, which also suggest a change in morphology with doping. The structural analyses of the samples are shown in Table. I. The lattice constants slightly increase with Ni doping so as grain size.

The effect of doping on the optical properties is examined by recording the absorption spectra of the Ni doped ZnO dispersions in ethanol from 250 to 550 nm (Fig. 2(a)). Two important observations can be seen from the figure - first is the redshift in the absorption edge and the second is the broadening of the excitonic peak, both with the increase in Ni concentration. Optical bandgap calculated from Tauc plot shown in the left inset of fig. 2(a) and right inset of fig.2 (a) shows optical band gap, which is decrease with increasing Ni concentration. To get a better understanding, we have calculated the optical bandgap from the plot of $(\alpha h\upsilon)^2$ vs $h\upsilon$ shown in the left inset of Fig. 2(a). It can be seen in the right inset in Fig. 2(a) that the optical bandgap decreases with increase in Ni concentration. The redshift in bandgap can be explained by considering the fact that Ni can easily replace the Zn from the ZnO lattice because $Ni^{2+}$ (0.69 Å) have smaller radii as compare to $Zn^{2+}$ (0.74 Å). the Ni ions have a radius of 0.69 Å, which is slightly lower than that of Zn (0.74 Å). As a result, Ni can quickly replace the Zn from the ZnO lattice. Therefore, the bandgap decreases due to the changes in the sp-d orbital electron exchange interactions produced by the localized d electrons of Ni in the matrix [13]. ZnO is characterized by a comparatively high exciton binding energy of 60 meV. The broadening of the excitonic peak with doping is attributed to the increase in dimension of the nanorods, which is clearly visible from the SEM. The optical absorption spectrum shows a weak absorption that extends into the long wavelength region (450-550 nm). This implies that the large size of the nanorods aggregates to cause the Mie effect, i.e. tails in the longer wavelength region of the absorption spectrum [18].

To further understand the role of Ni doping on the optical properties of the ZnO nanorods, we have shown in Fig. 2(b) the room temperature photoluminescence (PL). PL spectra of the samples display two distinct peaks. One is in the UV region corresponds to the excitonic emission and the second is a broad emission band in the visible region (450 - 600 nm) originating from the deep defects [19, 20]. It can be observed from the Fig. 2(b) that with the increase in the Ni concentration, the emission intensity in the UV region increases, which is attributed to the doping induced increase in the electron concentration [21]. The decrease in the visible emission is an indication of the reduction in the density of defects (such as oxygen vacancy $V_O$ and antisite oxide $O_{Zn}$) [22].

Open aperture Z-san which measures the total transmittance as a function of incident laser intensity was employed to study the off-resonant third-order nonlinear optical properties of the Ni doped ZnO nanorods. The Z-scan results of the samples at the peak on-axis intensity of 360

MW/cm$^2$ are presented in Fig. 3(a). As the bandgap of all the samples (~3 eV) are well above the single photon energy of 532 nm, it is reasonable to assume that a two photon process occurs in the samples. Expectedly, the Z-scan peak-shape response indicates the TPA for the samples. Importantly, the TPA shows a significant enhancement with the Ni concentration (Fig. 3(a)). To derive the TPA coefficient (β) of the samples, we have shown in Fig. 3(b) the intensity-dependent normalized transmittance, where the input intensity was tuned from 10 to 360 MW/cm$^2$. From the figure, it can be seen that, the transmittance decreases with increase in intensity that is consistent with the theory of TPA.

To get more detailed information on the variation of β with Ni doping in ZnO nanorods, we have fitted the experimental data with the Z-scan theory. In the case of TPA, the normalized transmittance as function of position z is given by [17]

$$T_N = \frac{1}{q_0\sqrt{\pi}} \int_{-\infty}^{+\infty} \ln(1 + q_0 e^{-t^2}) dt \qquad (1)$$

Where $q_0 = \frac{\beta I_0 L_{eff}}{\left(1 + z^2/z_0^2\right)}$ and $L_{eff} = \frac{(1-e^{-\alpha L})}{\alpha}$. $I_0$, $z_0$, L and α are the peak intensity at the focus (z=0), the Rayleigh length, sample thickness and linear absorption coefficient respectively. β values from the best-fit to the normalized transmittance are shown in the Table. II. It can be seen from the table that β value shows a significant enhancement from 7.6 ±0.4 to 112±6 cm/GW with Ni doping, i.e. 14 times increase when the Ni concentration is increased from 0 to 10%.

After demonstrating the substantial enhancement of the nonlinear optical response with Ni doping in ZnO nanorods, we have tried to explain the observed effects. For this, we have assumed that the increase of β with doping is due to the decrease in optical bandgap ($E_g$) of the sample. To get a consistent picture of this assumption, we have shown in Fig. 3(c), the variation of bandgap and β as a function of the doping concentration. It can be seen from the figure that β increases and $E_g$ decreases with increase in Ni concentration. Thus, we conclude that the large enhancement of β might be a result of bandgap reduction. Moreover, it has been theoretically and experimentally shown that the TPA in semiconductors follows an inverse cubic relation with bandgap [23], i.e. $\beta = \frac{K\sqrt{E_p}F}{n^2 E_g^3}$. In this equation, K and $E_p$ are material independent constants and *n* is the refractive index of the material. F is a band-structure dependent function, and its value depends on the ratio of photon energy and bandgap. For parabolic bands, F is given by the equation $F = \frac{[2h\nu/E_g - 1]^{3/2}}{[2h\nu/E_g]^5}$. From our experimental results, it can be seen that $E_g$ decreases with increase in Ni doping. Furthermore, a recent report [24] demonstrates that the refractive index decreases with Ni doping. Thus, the $n^2 E_g^3$ term decreases and consequently increases β. Additionally Ni (transition metal) doping leads to the redistribution of electronic charge and increases the polarizability of the medium. The enhancement in the third-order optical susceptibility ($\chi^3$) can be calculated from the rise in oscillator strength ($R^3/X_b^3$, where R is the crystallite size and $X_b$ is the exciton Bohr radius) [25]. From the Table. I, it can be seen that the R increases with metal doping, which in effect increases the β. Here we assumed that change in exciton Bohr radius is much smaller than the change in crystallite size with Ni doping. Expectedly, similar kind of enhancement in nonlinear absorption with metal doping is observed in semiconductor quantum dots [26]. For example, β shows 8 times enhancement with Cr doping and 11 times with Cu doping. Importantly, the nonlinear absorption and optical limiting properties of Ni doped ZnO nanorods are comparable or even better than other nanostructures such as liquid dispersion of carbon nanotubes (~50 MW/cm$^2$) [27], graphene nanostructures (~60 MW/cm$^2$) [28], and graphene/polyimide composites [29].

The remarkable observation of the substantial enhancement of TPA with Ni doping occurs at room temperature, is highly advantageous for potential applications in optical limiting and switching. The optical limiter is a nonlinear device that blocks the high-intensity light while being transparent to the lower intensity beams. They are very critical in protecting various optoelectronic detectors and eye from intense laser beams. In this context, we have shown in Fig. 3(d) a plot of output intensity ($I_{out}$) versus input intensity ($I_{in}$). From the figure, it can be seen that at low $I_{in}$, $I_{out}$ is linearly proportional to $I_{in}$ (Lambert-Beer law). As the $I_{in}$ increases a stage reaches where the $I_{out}$ is no longer linearly proportional to the $I_{in}$, i.e. deviation of the curve from the linear transmittance (the dashed line in Fig. 3(d)). The $I_{in}$ at which the curve deviates from the linear behavior is taken as the limiting threshold intensity that is an important performance parameter for an optical limiter. Further, it can be clearly seen from Fig. 3(c) that the limiting threshold intensity decreases with Ni doping, meaning optical limiting can be even achieved at moderate intensity with Ni doping. The variation of limiting threshold intensity with Ni ratio is shown in Table II.

In contrast to normal optical designing techniques where only two dimensional structuring restricted to the material surface is possible, TPA technique has the capacity to perform three dimensional manipulations. This is because the interaction of light with material takes place only at very high photon flux such as at the focusing point of a lens. By suitable laser focusing setups one can produce a well localized light matter interaction, precisely at any desired point inside the material. Thus, the high TPA in the doped samples brings many fold increase in its potentials and provides a way to engineer the material properties in three dimensions.

In conclusion, we have investigated the role of Ni doping on the third-order nonlinear optical properties of ZnO nanorods. In our experiments, we have demonstrated the substantial enhancement of the TPA coefficient as much as 14 times by the mere doping of 10% Ni. The significant increase in β values are correlated with the reduction in bandgap energy by Ni doping, which is in accordance with the bandgap scaling. Further we have also shown that the limiting threshold intensity decreases with doping, which make this material an excellent optical limiter. The improved TPA and optical limiting properties brings many fold increase in the technological potentials such as three dimensional patterning of this multifunctional material for the realization of various optoelectronic, magneto-optical and integrated devices.

**Acknowledgments**

This work was supported by the Department of Science and Technology (SERB-DST), India via a prestigious 'Ramanujan Fellowship' (SR/S2/RJN-121/2012) awarded to the PMS. PMS is thankful to Prof. Pradeep Mathur, Director, IIT Indore, for encouraging the research and providing the necessary facilities. A. J and K. V. A gratefully acknowledge the Department of Science and Technology (Project no: SR/S2/LOP-003/2010) and Council of Scientific and Industrial Research, India, (grant No. 03(1250)/12/EMR-II) for financial support.

TABLE I. Lattice parameters *a* and *c* of the Ni doped ZnO nanorods. Average grain sizes are calculated from the (002) reflection.

| Sample | *a* (nm) | *c* (nm) | $D_{(002)}$ (nm) |
|---|---|---|---|
| ZnO0 | 0.324 | 0.519 | 74.05 |
| ZnO5 | 0.324 | 0.520 | 76.69 |
| ZnO10 | 0.325 | 0.522 | 80.46 |

TABLE II. Experimentally calculated values of Eg, β and limiting threshold.

| Sample | $E_g$ (eV) | β (cm/GW) | Limiting Threshold (MW/cm$^2$) |
|---|---|---|---|
| ZnO0 | 3.01 | 7.6±0.4 | 132 |
| ZnO5 | 2.93 | 41.8±3 | 60 |
| ZnO10 | 2.85 | 112±6 | 30 |

**Figure captions**

**Fig. 1.** SEM images of (a) ZnO0, (b) ZnO5, (c) ZnO10 and (d) XRD of the samples.

**Fig. 2.** (a) Optical absorption spectra of the Ni doped ZnO nanorods dispersed in ethanol, left inset shows a representative curve for calculation of bandgap and the right inset represents the variation of bandgap with Ni concentration. (b) Photoluminescence spectra of the samples.

**Fig. 3.** (a) Open aperture Z scan curves of the samples. (b) Normalized transmittance as a function of input intensity. (c) Variation of β (left axis) and optical limiting threshold intensity (right axis) as function Ni doping (bottom axis) and bandgap (top axis). (d) Optical limiting curve for the samples and the dashed line shows the linear absorption regime. The solid line represents the theoretical fit.

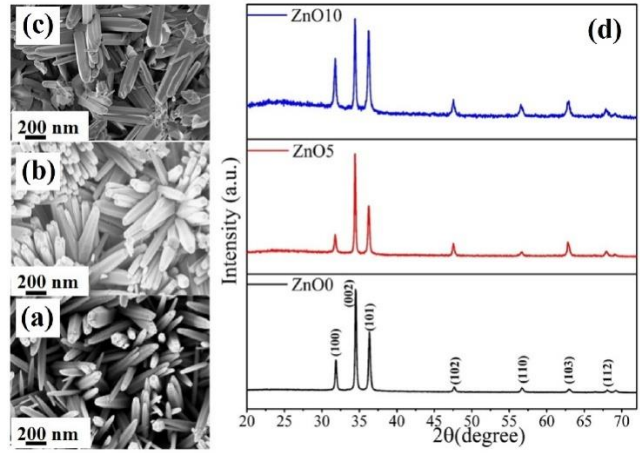

**Fig. 1.** SEM images of (a) ZnO0, (b) ZnO5, (c) ZnO10 and (d) XRD of the samples.

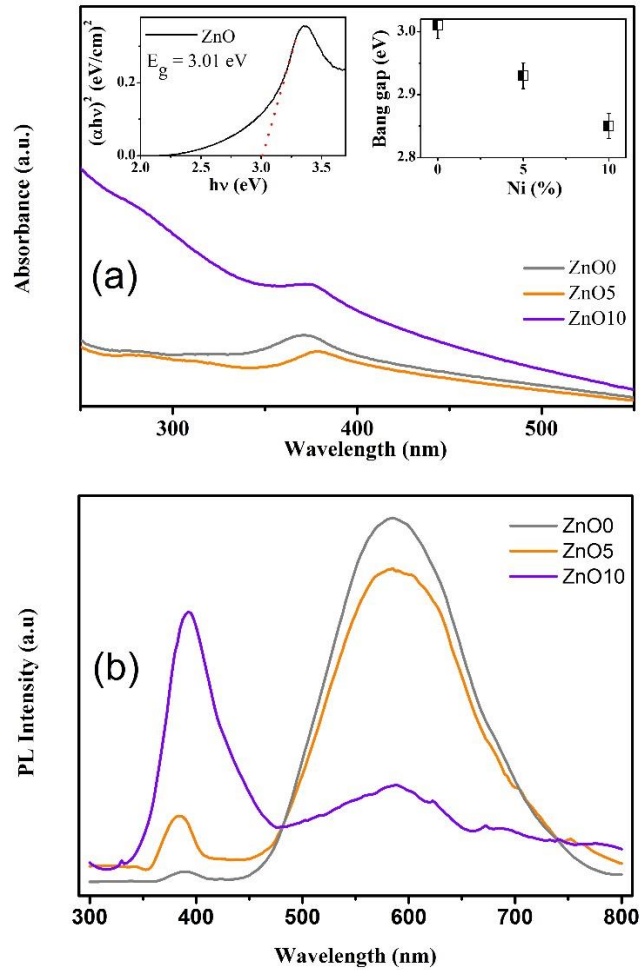

**Fig. 2.** (a) Optical absorption spectra of the Ni doped ZnO nanorods dispersed in ethanol, left inset shows a representative curve for calculation of bandgap and the right inset represents the variation of bandgap with Ni concentration. (b) Photoluminescence spectra of the samples.

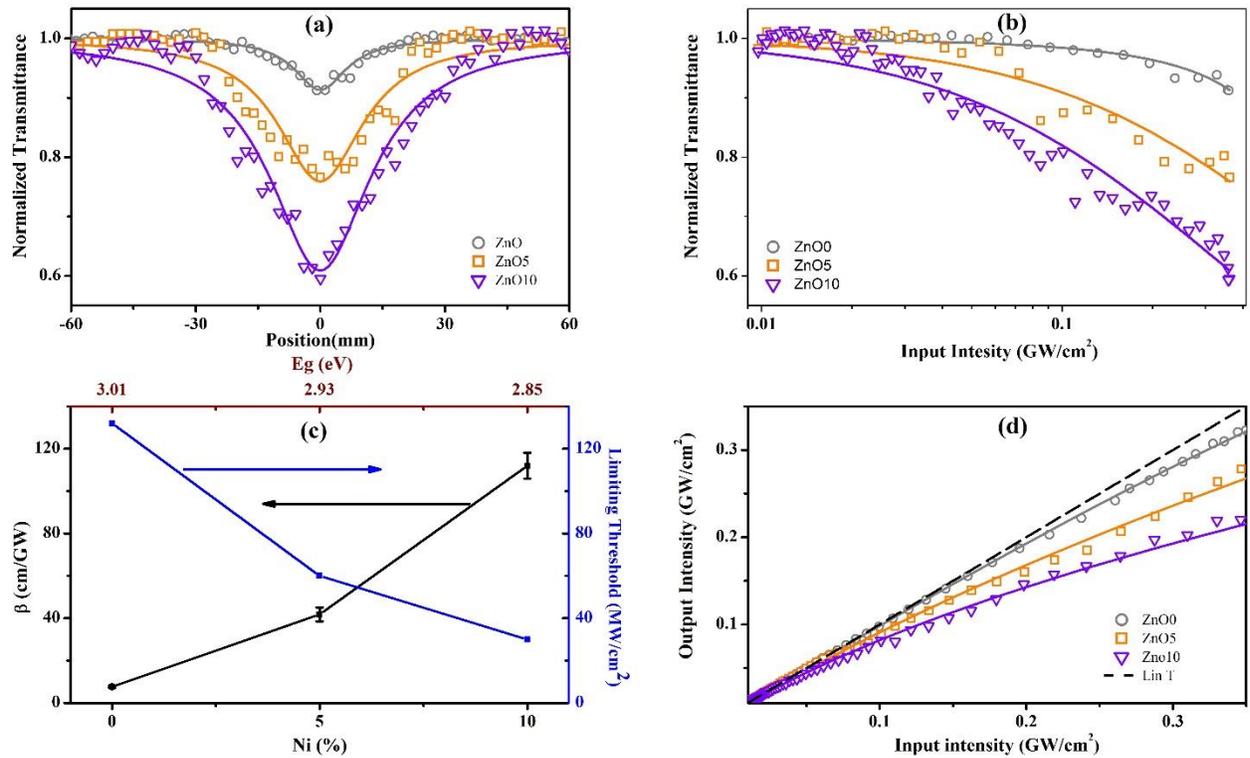

**Fig. 3.** (a) Open aperture Z scan curves of the samples. (b) Normalized transmittance as a function of input intensity. (c) Variation of β (left axis) and optical limiting threshold intensity (right axis) as function Ni doping (bottom axis) and bandgap (top axis). (d) Optical limiting curve for the samples and the dashed line shows the linear absorption regime. The solid line represents the theoretical fit.